\documentclass[aps, prl, reprint, amssymb, amsfonts, groupedaddress]{revtex4-1}

\pdfoutput=1

\usepackage{amsmath}
\usepackage{bm}
\usepackage{graphicx}
\usepackage[unicode]{hyperref}
\usepackage[latin1]{inputenc}
\usepackage{array}
\usepackage{upgreek}
\usepackage{color}

\begin{document}

\title{Density and $T_1$ of surface and bulk spins in diamond in high magnetic field gradients}

\author{M. de Wit} \thanks{Both authors contributed equally to this work}
\author{G. Welker} \thanks{Both authors contributed equally to this work}
\author{J. M. de Voogd}
\author{T. H. Oosterkamp}
\email{oosterkamp@physics.leidenuniv.nl}
	
\affiliation{Leiden Institute of Physics, Leiden University, PO Box 9504, 2300 RA Leiden, The Netherlands}

\date{\today}

\begin{abstract}
We report on surface and bulk spin density measurements of diamond, using ultra-sensitive magnetic force microscopy with magnetic field gradients up to 0.5 T/$\upmu$m. At temperatures between 25 and 800 mK, we measure the shifts in the resonance frequency and quality factor of a cantilever with a micromagnet attached to it. A recently developed theoretical analysis allows us to extract a surface spin density of 0.072 spins/nm$^2$ and a bulk spin density of 0.4 ppm from this data. In addition, we find an increase of the $T_1$ time of the surface spins in high magnetic field gradients due to the suppression of spin diffusion. Our technique is applicable to a variety of samples other than diamond, and could be of interest for several research fields where surface, interface or impurity bulk spin densities are an important factor.
\end{abstract}

\maketitle

\section{Introduction}
Noise coming from paramagnetic impurities is a widespread phenomenon, relevant to fields ranging from magnetometry to solid state qubits \cite{Bar-Gill2012, Ofori2012, Rondin2014}. For NV$^-$ centers in diamond (from on referred to as NV centers), the interaction with paramagnetic impurities is considered one of the main factors that induces decoherence of the NV center \cite{Bar-Gill2012}. This decoherence is faster for shallow NV centers close to the surface and slower for NV centers in the bulk of the diamond sample, because shallow NV centers are under the influence of a layer of electron spins at the surface of the diamond \cite{Myers2014, Ofori2012}. Understanding and potentially eliminating this source of decoherence has been a long-standing goal of the field \cite{Lee2017}. Here we present a new method to measure the impurity spin density, where the sensor is decoupled from the diamond sample. We use an ultrasoft cantilever with an attached micromagnet that couples to the spins via dipole-dipole interaction. The method is easily transferable to a wide range of samples \cite{Haan2015}. 

Multiple experiments have been conducted to measure the diamond surface impurity spin density and to characterize the properties of this two-dimensional electron spin bath, such as correlation times of the fluctuating spins \cite{Myers2014, Luan2015, Romach2015, Rosskopf2014, Grinolds2014}. The measured spin density values vary and range from $0.01$ to 0.5 $\mu_B$/nm$^2$. Most of these experiments are done at room temperature, except for one measurement at 10K \cite{Romach2015}. All mentioned studies use NV centers to probe the surface electron spin bath. The technological challenge of measuring surface or bulk spin densities on samples other than diamond can be met by using a scanning NV center approach \cite{Luan2015}. Unfortunately, the detection range of a scanning NV center is limited to a few nanometers. Our method is capable of sensing spins at micrometer distances.

We do our experiments at milliKelvin temperatures, where no surface spin density measurements on diamond have been performed yet. The low temperature in combination with a high magnetic field gradient allows us to measure with an extremely low force noise \cite{vinante2012}. In addition, it allows us to interact with electron spins that can easily be polarized by small magnetic fields and to disregard all physical processes involving phonons. This makes our method suitable for measuring spin densities in very dilute spin systems. In particular, it is of interest for the fields of quantum computation devices \cite{bruno2015,pappas2011} and magnetic resonance force microscopy (MRFM) \cite{poggio2010}, as surface and bulk impurity spins play an important role there.
 
Our group has previously demonstrated surface spin density measurements of dangling bonds on a silicon oxide surface \cite{Haan2015}. Here we present spin density measurements of paramagnetic impurities on a diamond surface and also expand our method to probe impurity spins in the bulk of the sample. We show that strong magnetic field gradients influence the $T_1$ relaxation time of the impurity spins and that this effect is an important ingredient to understand the system.

\section{Methods}
\subsection{Experimental setup}
\begin{figure}
\centering
\includegraphics[width=\columnwidth]{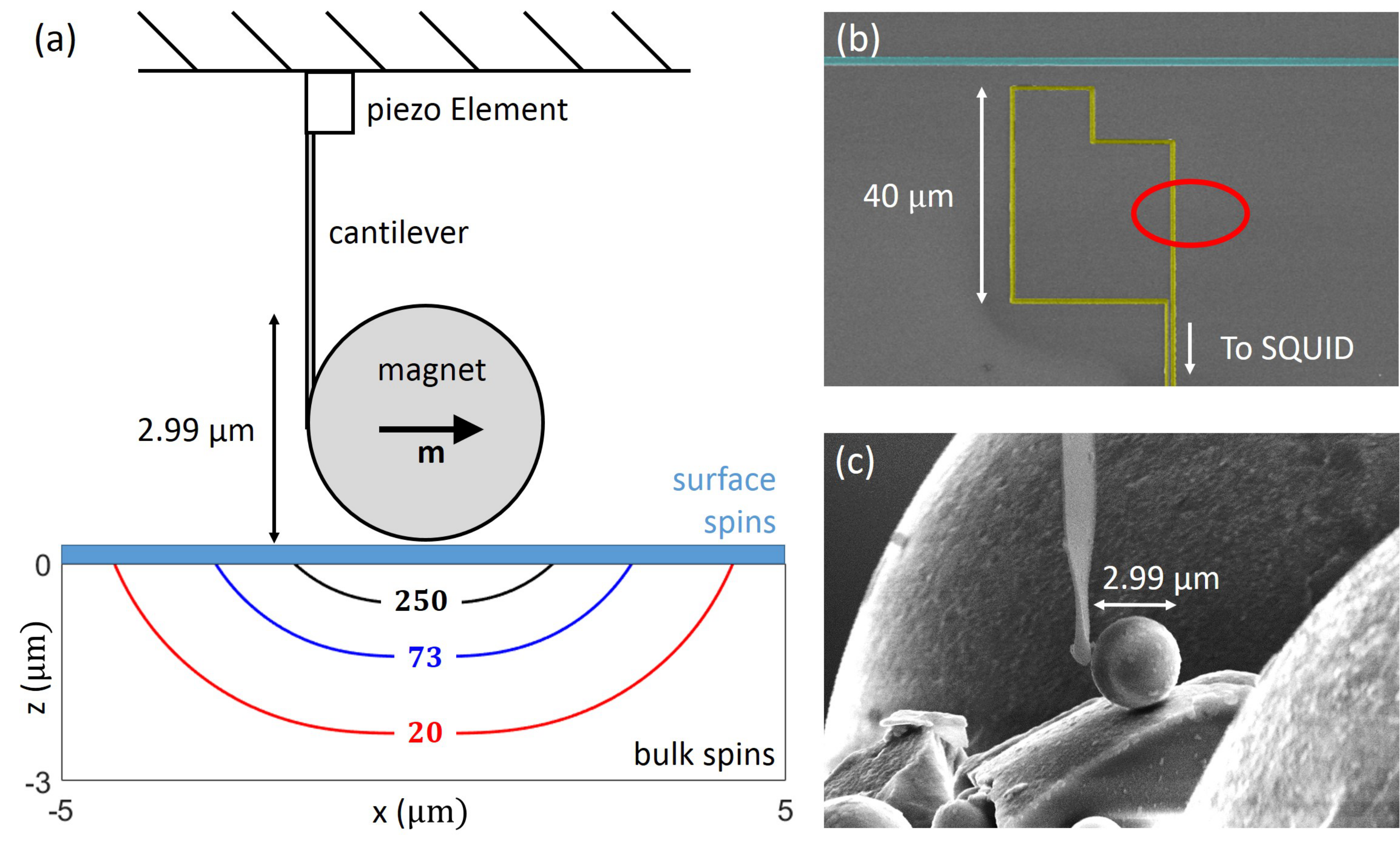}
\caption{(a) Setup: A magnetic particle with a diameter of 2.99 $\upmu$m attached to the end of a soft MRFM cantilever is positioned above the diamond sample, where it induces a high magnetic field gradient (units mT/$\upmu$m). The bulk of the diamond contains nitrogen impurities with an associated electron spin. On the surface we find an impurity layer containing paramagnetic electron spins, indicated in blue. (b) False colored scanning electron microscope image of the nanofabricated structures on top of the diamond sample. The pickup loop used for the readout of the cantilever is shown in yellow. In blue we see the NbTiN RF-wire, not used in the current experiment. The measurements described in this work were done at the location marked by the red circle. (c) Scanning electron microscope image of the tip of the cantilever and a NdFeB particle after the EBID.}
\label{figure:setup}
\end{figure}

In our experiments we use a commercially available diamond sample which has a size of 2.6 x 2.6 x 0.3 mm$^3$ and is specified to have less than 1 part per million (ppm) of nitrogen impurities \footnote{SC Plate CVD, \textless 100\textgreater, PL, from Element Six}. One surface is polished twice to an R$_{\mathrm{a}}$ \textless 5 nm \footnote{Second polish: scaife polishing from Stone Perfect}. We cleaned the diamond subsequently in acetone, 2-propanol, fuming nitric acid, hydrofluoric acid, and water in order to start the fabrication process with a clean surface and without oxides. On the surface we fabricated a niobium titanium nitride (NbTiN) pickup loop and RF-wire, the latter of which is not used in the present experiment \cite{thoen2017}. After fabrication, the sample was exposed to air for several months. Before mounting the sample, it was ultrasonically cleaned in acetone, and thereafter in 2-propanol to remove organics and dust.

The measurements were performed using an MRFM setup comparable to the one used in earlier experiments \cite{Haan2015}. To establish the magnetic interaction, we use a spherical NdFeB particle (from now on simply referred to as the magnet) with a diameter of 2.99 $\upmu$m. This magnet is glued with platinum using Electron Beam Induced Deposition (EBID) to the end of an ultrasoft cantilever with a length, width, and thickness of 166 $\upmu$m, 5 $\upmu$m, and 100 nm, respectively \cite{chui2003}. This geometry leads to a natural resonance frequency $f_n$ of 2850 Hz, and a spring constant $k_0 = m_{eff}(2 \pi f_n)^2 = 5.0 \times 10^{-5}$ N/m. After attaching the magnet, it is placed in an external field of 5 T, leading to a magnetic moment \textbf{m} of $1.5 \times 10^{-11}$ Am$^2$ pointing along the direction of movement of the cantilever (see Fig. \ref{figure:setup}(a)). The magnetic particle is responsible for the B-field that polarizes the spins in the sample, but also creates large magnetic field gradients of about 0.5 T/$\upmu$m.

The magnetized cantilever is mounted above the sample and can be moved with respect to the sample using a modified piezoknob-based cryogenic positioning stage \footnote{Cryo Positioning Stage - High Resonance (CPSHR), from Janssen Precision Engineering B.V.}. The absolute tip position is measured using three capacitive sensors, while the precise distance between the surface of the magnet and the surface of the diamond is calibrated by gently lowering the magnet until the two touch, using the piezoknobs.

\begin{figure}
\centering
\includegraphics[width=\columnwidth]{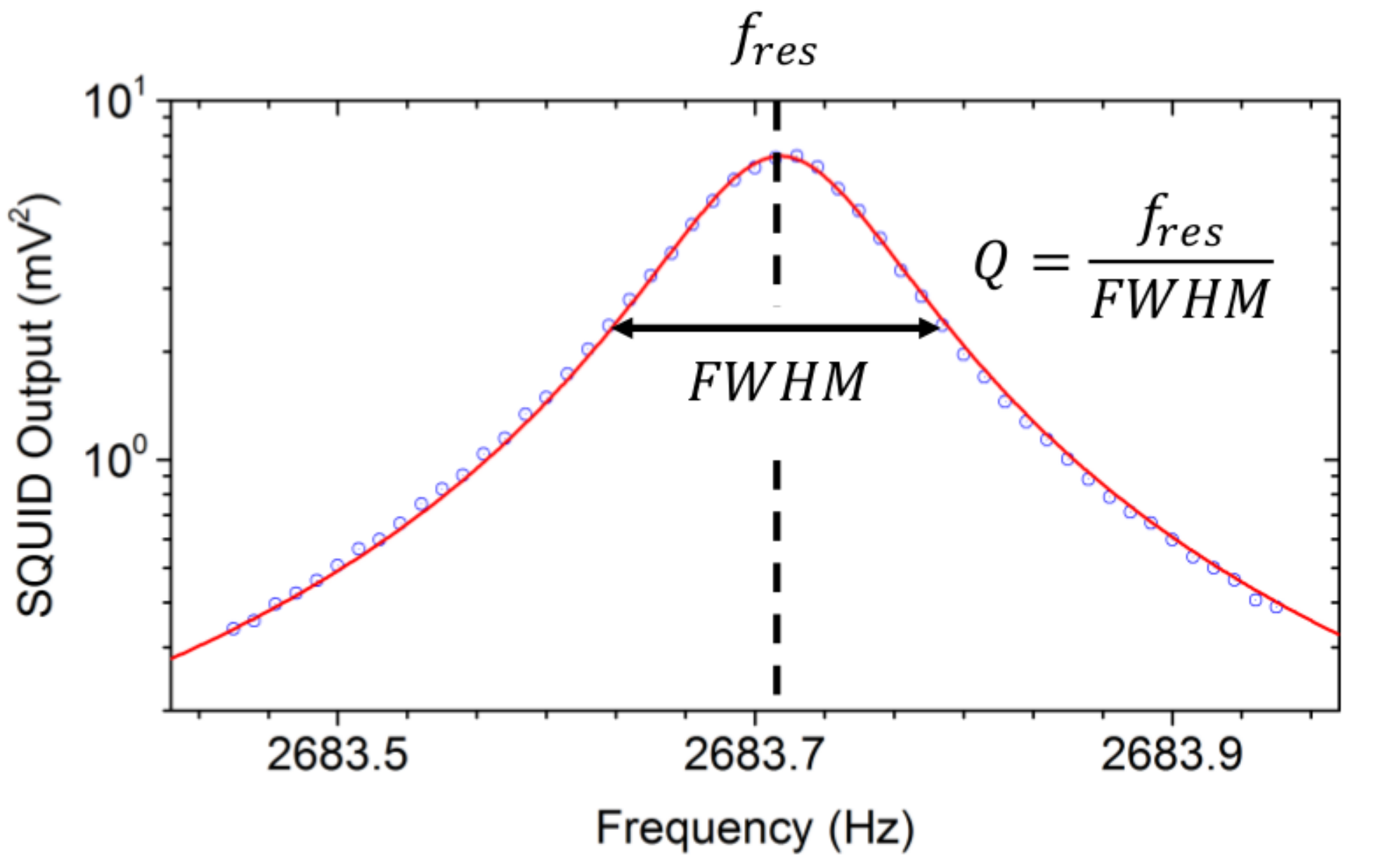}
\caption{Example of a frequency sweep measured at a tip-sample separation of 3.4 $\upmu$m at a temperature of 25 mK. The resonance frequency and quality factor are obtained by fitting the data to a Lorentzian (solid red line).}
\label{figure:sweep}
\end{figure}

The motion of the cantilever is measured using a SQUID-based readout \cite{usenko2011}, where we detect the changing magnetic flux in the pickup loop (colored yellow in Fig. \ref{figure:setup}(b)) due to the moving magnet. We can determine the linear response of the cantilever by driving a small piezo element at the base of the cantilever. When we sweep the drive frequency and measure the cantilever response using a lock-in amplifier, we obtain the resonance frequency and quality factor by fitting the square of the SQUID output with a Lorentzian, as seen in Fig. \ref{figure:sweep}.

The full experimental setup is mounted at the mixing chamber of a cryogen-free dilution refrigerator with vibration isolation \cite{haan2014}, and with a base temperature of 10 mK. The gold-plated copper sample holder is thermally connected to the mixing chamber using a silver strip. A heater and calibrated low temperature thermometer are used to control the temperature of the sample holder. Due to the limited thermal conductance between the mixing chamber, the sample holder, and the diamond sample itself, the sample temperature typically saturates at approximately 25 to 30 mK.

\subsection{Spin bath - cantilever coupling}
When the tip of the cantilever is positioned close to the sample, it couples to the electron spins via the magnetic field that originates from the magnet. This coupling results in a shift $\Delta f$ of the resonance frequency $f_{res} = f_0 + \Delta f$, with $f_0$ the resonance frequency without coupling to the spin bath, and an increase in the dissipation which can be seen as a shift in the inverse quality factor $\Delta \frac{1}{Q}$. This interaction was recently investigated by De Voogd \textit{et al.} following a Lagrangian approach taking into account the full dynamics of the spins \cite{voogd2017}.

In our sample, we expect two main sources for the signal. First of all, we expect a contribution from the free electron spins associated to the nitrogen impurities in the bulk of diamond (P1 and P2 centers). Because of the long $T_1$ relaxation times of the dilute electron spins in the bulk, which have been reported to increase to several seconds at low temperatures \cite{takahashi2008}, we expect the bulk-induced shift of the quality factor to be zero, resulting in a final contribution given by
\begin{align}
\Delta f_{bulk} &= \frac{f_0}{2k_0} \frac{\rho \mu_B^2}{k_B T} \int_{V} d^3 \bm{r} ~ \mathcal{C}(\bm{r}) \text{,}\qquad\text{and} \\ \label{eq:fbulk}
\Delta \frac{1}{Q}_{bulk} &= 0 \text{,}\qquad\text{where} \\
\mathcal{C}(\bm{r}) &= \frac{|\mathbf{B'}_{||\mathbf{\hat{B}_0}}|^2}{\cosh^2 (\frac{\mu_e B_0}{k_B T})}, \label{eq:C}
\end{align}
and $\rho$ the bulk spin density, $\mu_B$ the Bohr magneton, and $T$ the temperature of the spin bath.

Since our sample was exposed to air before the experiment, we expect a second contribution from a layer of surface spins which can be expected on any surface which has been exposed to the air for extended times \cite{peddibhotla2013}. Based on our earlier experience with the surface spins on the surface of silicon, we expect these spins to have $T_1$ times similar to $\frac{1}{\omega_0}$. Therefore, these spins should cause additional shifts given by
\begin{equation} \label{eq:fsurf}
\Delta f_{surf} = \frac{f_0}{2k_0}  \frac{\sigma \mu_B^2}{k_B T} \int_{S} d^2 \bm{r} ~ \mathcal{C}(\bm{r}) \frac{(\omega_0 T_1(\bm{r}))^2}{1+(\omega_0 T_1(\bm{r}))^2},
\end{equation}
and
\begin{equation} \label{eq:qsurf}
\Delta \frac{1}{Q}_{surf} = \frac{1}{k_0} \frac{\sigma \mu_B^2}{k_B T} \int_{S} d^2 \bm{r} ~ \mathcal{C}(\bm{r}) \frac{\omega_0 T_1(\bm{r})}{1+(\omega_0 T_1(\bm{r}))^2},
\end{equation}
where $\omega_0 = 2 \pi f_0$, and $\sigma$ is the surface spin density. Please note that we have placed the term containing $\omega_0 T_1$ inside the integral to reflect the fact that $T_1$ may depend on the magnetic field gradient.

In order to calculate the expected frequency shift and additional dissipation, accurate values are needed for the magnetic moment, shape and size of the magnetic field. In our experiment, since the magnetic particle is almost perfectly spherical, we can calculate the field as if it originates from a magnetic dipole. In the coordinate-free form, this is given by \cite{Griffiths1962}:
\begin{equation} \label{eq:Bdip}
\mathbf{B}(\mathbf{r}) = \frac{\mu_0}{4 \pi} \frac{1}{r^3} \left[ 3 \left( \mathbf{m} \cdot \mathbf{\hat{r}} \right) \mathbf{\hat{r}} - \mathbf{m} \right],
\end{equation} 
with \textbf{m} the magnetic moment of the magnet. From this field, we can calculate all relevant derivatives as required for Eq. \ref{eq:C}.

\subsection{Spin diffusion in high magnetic field gradients}
The theory presented so far describes the spin-cantilever interaction for a constant $T_1$ of the spins. For most applications, for instance in bulk techniques with homogeneous external fields, this is a good approximation. However, this approximation does not hold when dilute spins are placed in large magnetic field gradients, as is the case in our experiment. These gradients can increase the relaxation times by suppressing spin diffusion, a concept first derived by Bloembergen \cite{bloembergen1949}. Spin diffusion in diamond was studied before by Hammel \textit{et al.} \cite{cardellino2014}.

In this model, it is assumed that different spins can have different relaxation times based on their local environment. This results in the presence of fast-relaxing spins which can rapidly thermalize to the lattice, and slow-relaxing spins which are badly coupled to the lattice. After a perturbation of the thermal equilibrium, relaxation of the polarization of this sample back to equilibrium occurs via spin diffusion which couples the slower relaxing spins to the faster relaxing spins through flip-flop interaction, reducing the overall relaxation time of the sample.

However, spin diffusion can be suppressed by applying a large magnetic field gradient, which reduces the probability of two spins exchanging energy by introducing a difference in field felt by neighbouring spins. An Ansatz for the suppression of the spin diffusion can be obtained by calculating the normalized overlap interal between the lineshapes of two spins \cite{budakian2004}:
\begin{equation} \label{eq:Overlap}
\Phi(G) = \frac{\int f(B')f(B'-\bar{z}G) dB'}{\int f^2(B') dB'},
\end{equation}
with G the gradient of the magnetic field strength at the position of the spins, $\bar{z} = \bar{r}/2$ the average spacing between spins, and $f(B)$ the resonance lineshape of the spins. Since we are considering a layer of spins on the surface of the diamond, the total number of spins is too small to measure the actual spectra of the surface spins using bulk techniques like ESR, so we assume these spins to have a Lorentzian profile with a linewidth given by \cite[p.~128]{abragam1961}:
\begin{equation} \label{eq:linewidth}
\Delta B_{dd} = 3.8 \mu_0 \gamma_e \hbar / 4 \pi \bar{r},
\end{equation}
with $\gamma_e/2\pi =$ 28.0 GHz/T the electron gyromagnetic ratio.

Since the correlation function of two Lorentzian profiles with width $\Delta B_{dd}$ is itself a Lorentzian with a width twice as large, we find that the relaxation time is given by
\begin{equation} \label{eq:T1_vs_Grad}
T_1(G) = \left[ \frac{1}{T_1^{ff}} \frac{1}{(1 + (G/G^*)^2)} + \frac{1}{T_1^*} \right]^{-1},
\end{equation}
with $T_1^{ff}$ the reduced $T_1$ time due to flip-flops between neighboring spins, and $T_1^*$ the intrinsic relaxation time of the system when the flip-flops are completely quenched. $G^*$ is a measure for the gradient when the quenching becomes significant, from now on called the critical gradient, and can be determined by calculating when the difference in field at neighboring spins becomes larger than the spin linewidth, i.e., when $\bar{z} G^* > \Delta B_{dd}$. Note that this is only a heuristic description of effects of spin diffusion, as we do not take into account the direction of the gradient, nor the effects of the spin bath polarization on the flip-flop rate.

\begin{figure}
\centering
\includegraphics[width=\columnwidth]{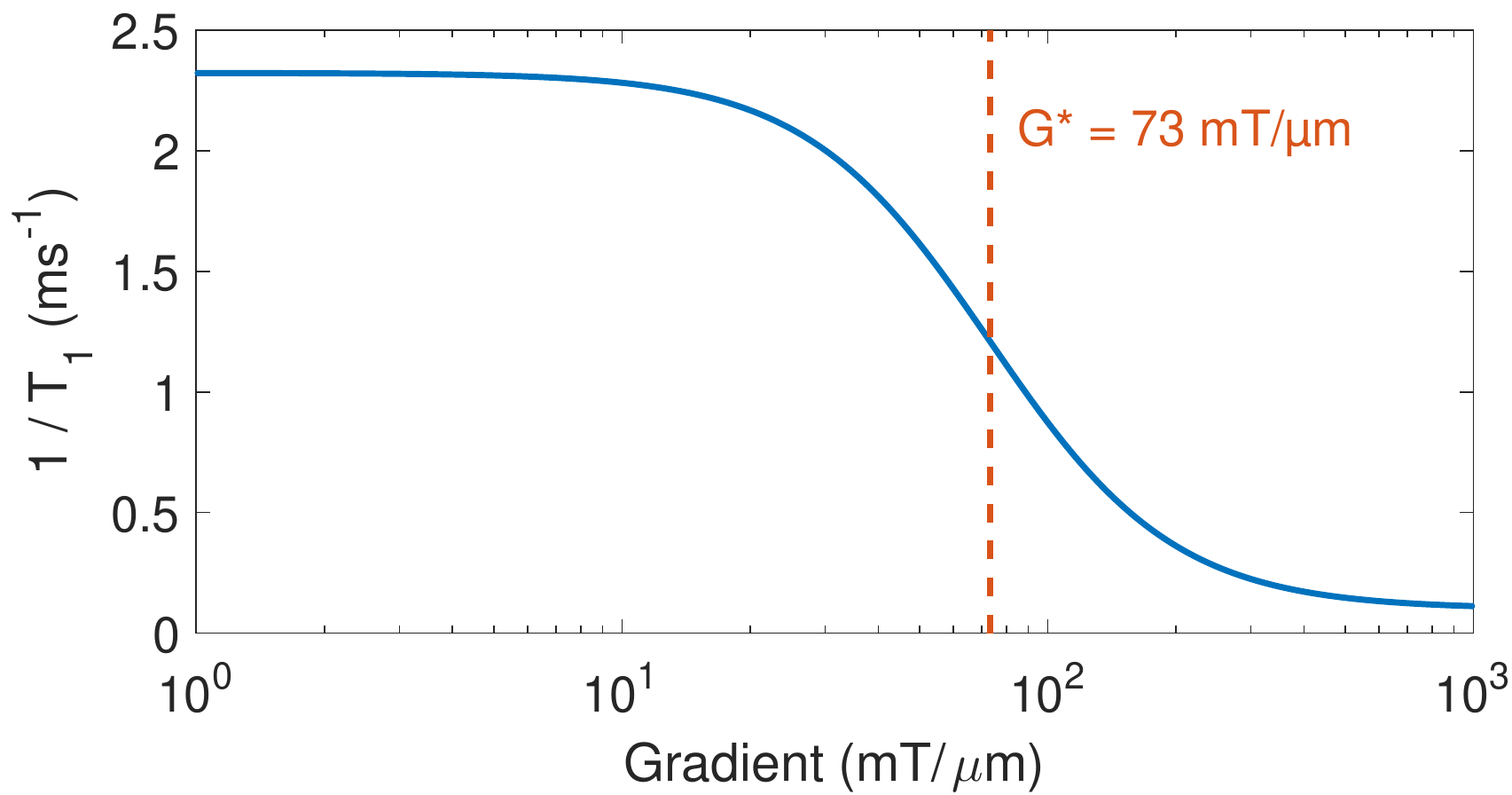}
\caption{Plot of the inverse of the $T_1$ time calculated from Eq. \ref{eq:T1_vs_Grad}, using $G^*$ = 73 mT/$\upmu$m, $T_1^{ff}$ = 0.45 ms, and $T_1^*$ = 10 ms.}
\label{figure:T1vsGrad}
\end{figure}

\section{Results and discussion}
For the experiment, we use the aforementioned positioning system to position the magnet at certain heights above the sample. The height is defined as the distance between the surface of the diamond, and the surface of the magnet. At each height the temperature is varied from 25 mK up to 800 mK. At every height-temperature combination, the resonance frequency and quality factor are measured as described in Sec. \emph{Experimental setup}.

\subsection{Frequency shift and dissipation}
\begin{figure}
\centering
\includegraphics[width=\columnwidth]{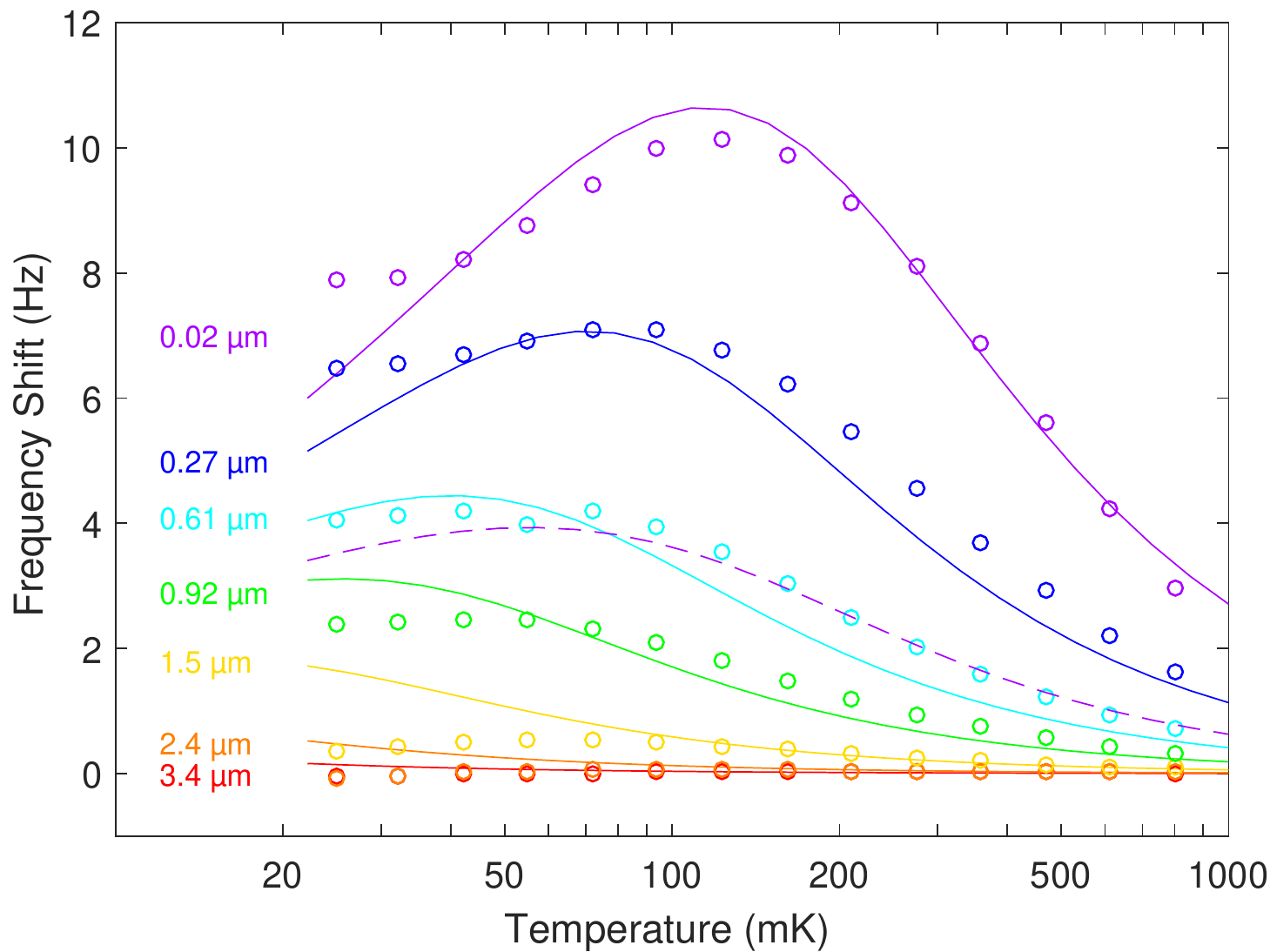}
\caption{Data (circles) and theory (lines) for the frequency shift of the cantilever versus temperature when positioned near the surface of the diamond sample. The dashed line shows the contribution from the bulk spins in the diamond only. The solid lines were calculated using $\sigma = 0.072$ spins/nm$^2$, and $\rho = 0.40$ ppm.}
\label{figure:MFM_df}
\end{figure}

The results of the measurements of the frequency shift are shown in Fig. \ref{figure:MFM_df}. The solid lines indicate the results of the fits according to Eq. \ref{eq:fbulk} and Eq. \ref{eq:fsurf}, with the total frequency shift given by $\Delta f = \Delta f_{bulk} + \Delta f_{surf}$. We calculated $f_0$ at each height by extrapolating the measured frequency shift data to higher temperatures.

The 2D and 3D integrals over $\mathcal{C}$ are calculated using the magnetic field distribution defined by Eq. \ref{eq:Bdip}. The only free parameters remaining in the model are the two spin densities $\rho$ and $\sigma$ for the bulk and surface, respectively, and the $T_1$ time of the surface spins, which for now is fixed at a value of 0.5 ms. As the term $\frac{(\omega_0 T_1)^2}{1+(\omega_0 T_1)^2}$ converges to 1 for $\omega_0 T_1 \gg 1$, the effect of the $T_1$ time on the total frequency shift can be neglected, so the precise value for the $T_1$ time only becomes important in the analysis of the temperature dependent change of the quality factor.

A complication in fitting the values for the two spin densities, is that the functions for $\Delta f_{bulk}$ and $\Delta f_{surf}$ are not independent. To determine the precise values, we fixed $\rho$, and fitted $\sigma$ over the temperature traces for each height. Next, we varied $\rho$ to minimize the average fitting error. This method yields global values of {$\rho$ = 0.4 ppm}, compatible with the specifications of the diamond sample, and $\sigma$ = 0.072 spins/nm$^2$. The dashed line in Fig. \ref{figure:MFM_df} shows the frequency shift due to the bulk spins at a height of 20 nm for this concentration, signifying that even very low spin densities have a substantial effect on the total frequency shift.
 
\begin{figure}
\centering
\includegraphics[width=\columnwidth]{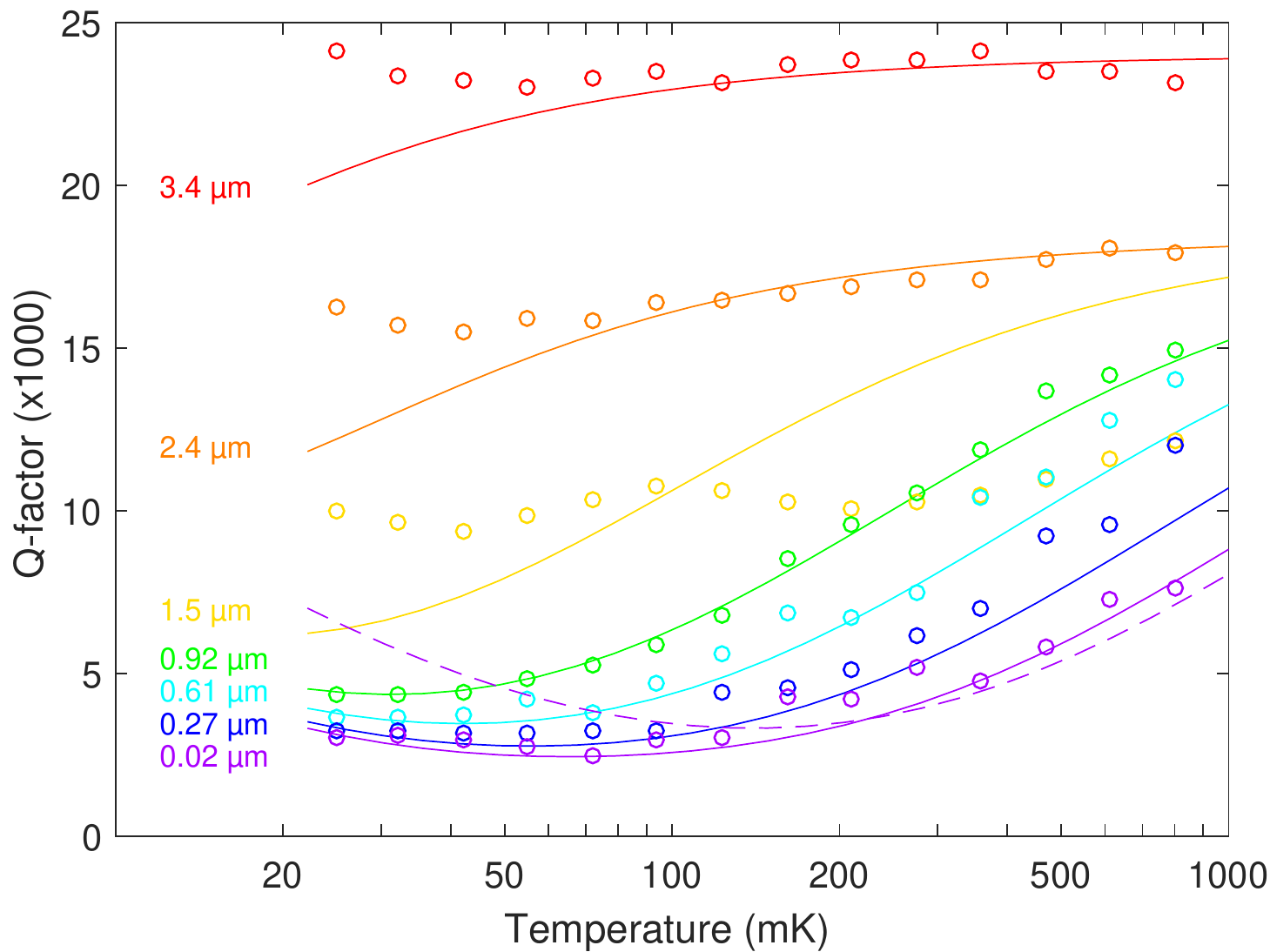}
\caption{Data (circles) and theory (lines) for the change in the quality factor of the cantilever versus temperature when positioned near the surface of the diamond sample. The solid lines are calculated using the spin densities obtained from the frequency data, including the effects of spin diffusion using $T_1^* = 10$ ms and $T_1^{ff} = 0.45$ ms. The dashed line shows the expected quality factor at a height of 0.02 $\upmu$m calculated using a constant $T_1 = 1.3$ ms.}
\label{figure:MFM_dQ}
\end{figure}

The measured changes of the quality factor for each height and temperature are shown in Fig. \ref{figure:MFM_dQ}. The total value for the inverse quality-factor is given by $\frac{1}{Q} = \frac{1}{Q_0} + \Delta \frac{1}{Q}_{surf}$, with $Q_0$ the quality factor of the resonator without coupling to the spin bath. For large heights, we obtain $Q_0$ by extrapolating the measured dissipation to high temperatures. For small heights, we set it to a constant value of 18500. This value is much lower than the vacuum quality factor of about 50000, probably due to some other long-ranged effect, for instance electrostatic interactions \cite{stipe2001,kuehn2006}.

To fit this data to Eq. \ref{eq:qsurf}, we fixed the spin densities of both the surface and the bulk to the values obtained from the frequency shift analysis. Trying to fit this data solely using Eq. \ref{eq:qsurf} did not yield a good match with the data, as illustrated by the dashed line in Fig. \ref{figure:MFM_dQ}, which shows the result of the calculation at a height of 20 nm, with $T_1 = 1.3$ ms. A clear deviation between the data and calculation at low temperatures is visible. Repeating this approach for all available heights results in a clearly increasing $T_1$ time for smaller tip-sample separations. This observation is a strong indication for the suppression of the spin diffusion by the high magnetic field gradient.

We have included this effect by inserting Eq. \ref{eq:T1_vs_Grad} into Eq. \ref{eq:qsurf}, yielding a position-dependent $T_1$ time bound by $T_1^*$ in the high gradients close to the magnet, and $T_1^{ff}$ for spins far away from the magnet. Using the surface spin density $\sigma = 0.072$ spins/nm$^2$ obtained from the frequency shift data, we find that in our case $\bar{r} = \sigma^{-1/2} = 3.7$ nm, resulting in a linewidth of $\Delta B = 0.14$ mT according to Eq. \ref{eq:linewidth}. This leads to a critical gradient $G^* = 73$ mT/$\upmu$m, a value smaller than the maximum field gradients in our setup as indicated in Fig. \ref{figure:setup}(a). The resulting dependence of the $T_1$ time as a function of the magnetic field strength gradient is shown in Fig. \ref{figure:T1vsGrad}.

\begin{figure}
\centering
\includegraphics[width=\columnwidth]{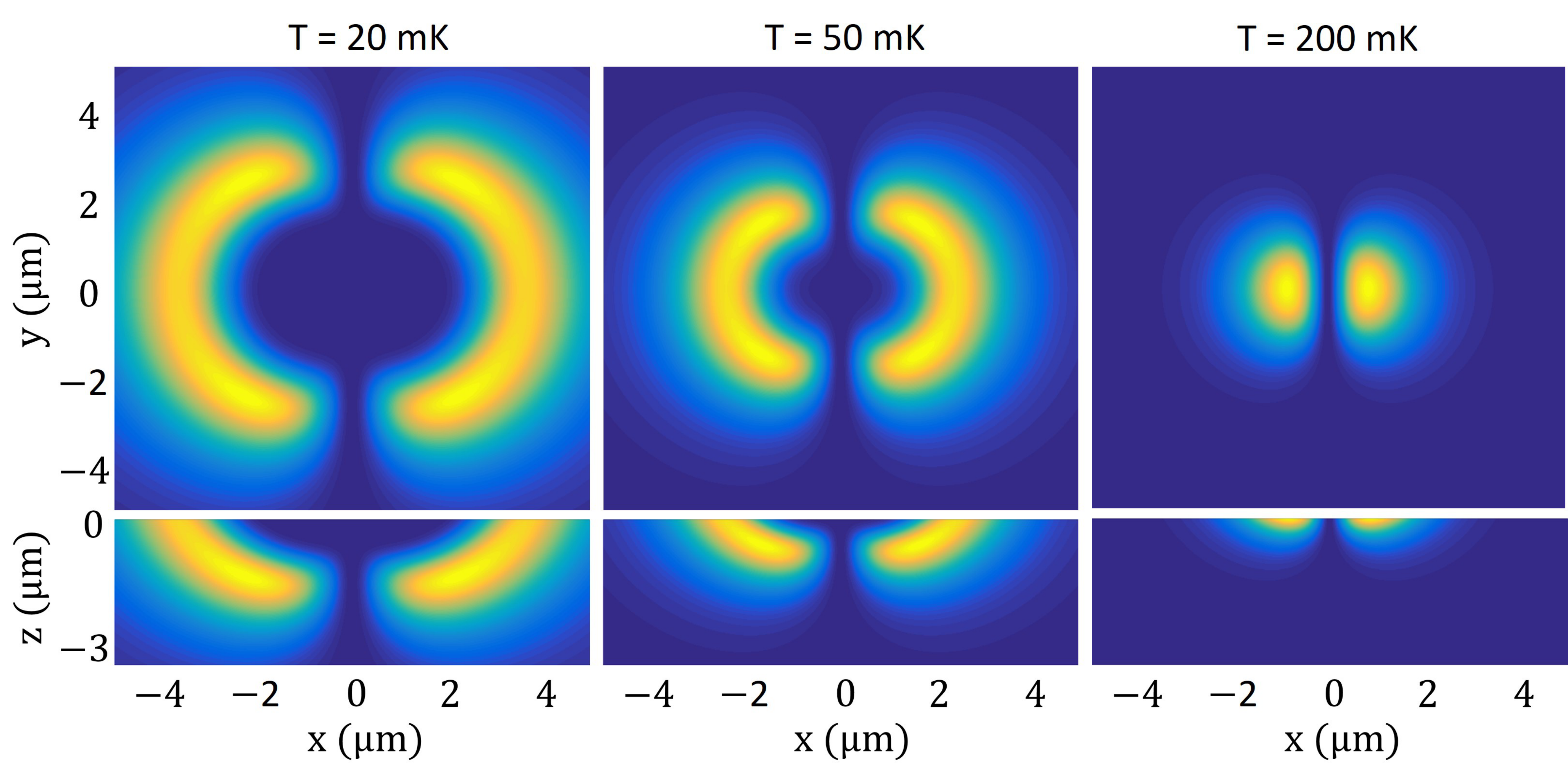}
\caption{Simulation of the relative contribution of spins at different locations, calculated for a tip-sample separation of 20 nm. Yellow indicates regions of maximal coupling, while blue indicates a very low coupling between a spin and the cantilever.}
\label{figure:contribution}
\end{figure}

To obtain reliable values for the two relaxation times $T_1^{ff}$ and $T_1^*$, we make use of an interesting feature of the coupling between the spins and the magnet. Fig. \ref{figure:contribution} shows the spatial distribution of $\mathcal{C}$ for various temperatures, calculated at a constant tip-sample separation of 20 nm, indicating the position of the spins contributing most to the signal. It is clear that as the temperature of the sample decreases, the average location of contributing spins moves away from the cantilever. This immediately implies that at low temperatures, most of the contributing spins are located in a region with a magnetic field gradient below $G^*$, which means that spin diffusion is not suppressed, and thus their relaxation time approaches $T_1^{ff}$. Equivalently, at high temperatures, the spins that contribute the most are close to the magnet in a high magnetic field gradient, meaning flip-flops are quenched and $T_1 \approx T_1^*$. This allows us to fit $T_1^{ff}$ and $T_1^*$ almost independently. The solid lines in Fig. \ref{figure:MFM_dQ} show the final calculations including the effects of spin diffusion using $T_1^* = 10$ ms and $T_1^{ff} = 0.45$ ms.

We selected a value of 10 ms for $T_1^*$. Higher values for $T_1^*$ do not significantly change the dissipation, because $T_1^*$ gets too far away from the cantilever period. In other words: the dissipation of the cantilever mediated by the spins peaks when $T_1$ matches the cantilever period, so we are only sensitive to $T_1$ times of up to several milliseconds. Spins with a $T_1$ time larger than several milliseconds do not contribute to the enhanced dissipation, but they do change the resonance frequency.

There are still some unexplained features in the data. First of all, there is a clear difference between data and theory for the large tip-surface separations at low temperatures. It seems that the quality factor of the silicon cantilever increases when the temperature decreases, presumably due to the freezing out of the quantum fluctuators on the surface of the silicon beam \cite{vinante2017}. Furthermore, the measurements at a height of 1.5 $\upmu$m also strongly deviate from the fit for both the resonance frequency and the quality factor. This could be due to the fact that this measurement was performed directly above a superconducting line of the pickup loop, which might lead to a lower density of paramagnetic electron spins on and beneath the superconductor. The low quality factor can then be explained by the increased coupling with the pickup loop, which leads to additional dissipation of the cantilever energy via the inductive coupling to resistive elements. We did not take the data measured at this height into account in our final analysis.

\section{Summary and outlook}
In conclusion, by using our MRFM setup as an ultrasensitive, long-range MFM, we have been able to measure the amount of nitrogen impurities in our diamond sample, resulting in a bulk spin density of only 0.4 ppm. This shows that our method allows us to characterize samples containing low spin densities over a field of view of several micrometers. Furthermore, we have characterized the paramagnetic electron-like spins on the surface of the diamond, yielding a density of 0.072 spins/nm$^2$, and $T_1$ times of several milliseconds, heavily influenced by the presence of spin diffusion. As it is the fluctuation of these spins that is typically held responsible for the reduced performance of a variety of nanodevices like qubits and superconducting resonators, we believe that our technique offers a useful tool to characterize the properties of the surface spin system, and understand the resulting dissipation in these devices.

As the flip-flop interaction between the surface spins on the diamond can be reduced by using a high gradient, it could be possible to improve the coherence of various diamond-based devices. The idea of suppressing flip-flop induced spin bath fluctuations for this purpose has been demonstrated before by increasing the polarization of the spin bath to over 99\% \cite{takahashi2008}. However, this only works for low temperatures and high magnetic fields, and is very challenging for nuclei due to the small magnetic moment. These drawbacks do not apply for gradient-based quenching of flip-flops. Furthermore, since the required magnitude of the critical gradient depends on the spin density, relatively modest magnetic field strenght gradients are required to isolate a single spin from its environment in very pure samples. For example, to suppress spin diffusion in a diamond sample with a nitrogen spin density of 1 ppm, it is sufficient to have a gradient of 1 mT/$\upmu$m.

A potential near-future application of this technique could be the testing of various sample preparation steps that are typically used in order to enhance the performance of nano\-de\-vices. As an example, we expect that a short chemical wet etch of the diamond using hydrofluoric acid should reduce the density of the unpaired spins on the surface, resulting in the case of MRFM in a higher quality factor of the resonator close to the surface, and in the case of shallow NV centers in enhanced correlation times. Our technique would allow us to test the effect of this etch in any intermediate state of the fabrication of one of these devices, allowing for a better optimization of the fabrication process.

\section{Acknowledgments}
We thank F. Schenkel, J. P. Koning, G. Koning and M. Camp for technical support. We thank D. J. Thoen, T. M. Klapwijk, and A. Endo for providing us with the NbTiN. We thank T. van der Sar for valuable discussions. We thank J. J. T. Wagenaar for proofreading this manuscript.  This work is part of the single phonon nanomechanics project of the Foundation for Fundamental Research on Matter (FOM), which is part of the Netherlands Organisation for Scientific Research (NWO).

\section{Author Contributions}
J.d.V. designed and built the setup and fabricated the sample. M.d.W, G.W. and J.d.V. performed the measurements. J.d.V. and M.d.W did the analysis of the data. M.d.W and G.W. wrote the manuscript. All authors discussed the results and reviewed the manuscript. M.d.W, G.W. and J.d.V. have equal contribution in this work.

\bibliography{Spin_Density_Diamond_Literature}

\end{document}